\documentclass[showpacs,showkeys,preprintnumbers]{revtex4-1}
\usepackage{amssymb}

\usepackage{amsfonts}

\usepackage{latexsym}

\usepackage{dcolumn}

\usepackage{amsmath}

\usepackage{graphicx}

\usepackage{psfrag,pstricks}

\newcommand{\eqn}{\ref}

\begin{document}

\title{Dynamical Behaviours of the Nonlinear Atom-Field Interaction in the Presence of Classical Gravity: \textit{f}-Deformation Approach}

\author{ Sh. Barzanjeh$^{1}$ }
\email{shabirbarzanjeh@yahoo.com}

\author{M. H. Naderi$^{2}$}
\email{mhnaderi@phys.ui.ac.ir}
\author{ M. Soltanolkotabi$^{2}$}
\email{soltan@sci.ui.ac.ir}
\affiliation{ $^{1}$ Department of Physics, Faculty of Science, University of Isfahan, Hezar Jerib, 81746-73441, Isfahan, Iran\\
$^{2}$Quantum Optics Group, Department of Physics, Faculty of Science, University of Isfahan, Hezar Jerib, 81746-73441, Isfahan, Iran}

\date{\today}

\begin{abstract}

\textbf{Abstract}\\
In this paper, we investigate the effects of a classical gravitational field on
the dynamical behaviour of nonlinear atom-field interaction within the framework of the \textit{f}-deformed Jaynes-Cummings
model. For this purpose, we first introduce a set of new atomic operators obeying an \textit{f}-deformed \textit{su(2)} algebraic structure to
derive an effective Hamiltonian for the system under consideration. Then by solving the Schr\"{o}dinger equation in the
interaction picture and considering certain initial quantum states for the atomic and radiation subsystems, we analyze the
influence of gravity  on the temporal evolution of the atomic population inversion, atomic dipole squeezing, atomic momentum diffusion,  photon counting statistics, and deformed quadrature squeezing of the radiation field.

\end{abstract}

\maketitle

%TEXT--------------------------------------------------------------------------
\section{Introduction}

The quantum electrodynamical model involving the one-photon interaction of a single quantized mode of the electromagnetic field with a motionless two-level atom, known popularly as the Jaynes-Cummings (JC) model~\cite{1} in the literature, has been extensively studied in the last four decades. The JC model has predicted many unexpected and non-trivial results, e.g. the well known phenomenon of collapses and revivals of Rabi oscillations- unquestionable evidence of the quantum nature of the electromagnetic field~\cite{2}. The other non-trivial and fascinating quantum features displayed by the JC model are squeezing of the cavity-field~\cite{3}, atomic dipole squeezing~\cite{4} and atom-cavity entanglement~\cite{7}. At present, the JC model occupies a special place in quantum optics because it helps to examine and verify the conjectures about more complicated models which are close to real processes. Recent experiments with Rydberg atoms and microwave photons in a superconducting cavity have turned the JC model from a theoretical curiosity to a useful and testable enterprise~\cite{8}. Such a system is also suitable for quantum-state engineering and quantum information processing. Up to now, Fock states~\cite{9}, Schr\"{o}dinger cat states~\cite{10}, and entangled states~\cite{11} have been produced in cavity quantum electrodynamics experiments.

Stimulated by the success of the (standard) JC model, it has been generalized and extended in many interesting directions to explore new quantum effects. Discussions related to several interesting generalizations of this model are now available in the literature~\cite{13,14,16,18,19,20,25,38}. Among the generalized versions of the JC model the so-called \textit{ f}-deformed JC model (see, e.g. Refs.~\cite{16, 19, 25,38, 41, 42, 44}, and~\cite{46}) has received much attention in view of its connection with quantum algebras~\cite{48}. The quantum algebras, which appear as a generalization of the symmetry concept~\cite{49} and the basics of so-called non-commutative theories, have given the possibility of generalizing the notion of creation and annihilation operators of the usual quantum oscillator and to introduce deformed oscillator. Some deformed versions of oscillator algebra have found many applications to various physical problems, such as the treatment of vibrational spectra of molecules~\cite{50} and the investigation of nonlinearities in quantum optics~\cite{51}. In addition, it has been shown~\cite{25} that most of the nonlinear generalizations of the JC model are only particular cases of the \textit{f}-deformed JC model in which the creation and annihilation operators of the radiation field are replaced by deformed harmonic-oscillator operators with prescribed commutation relations. The representation theory of the quantum algebras with a single deformation parameter \textit{q} has led to the development of the \textit{q}-deformed oscillator algebra~\cite{52}. By using a \textit{q}-oscillator description, the JC model Hamiltonian with an intensity-dependent coupling has been generalized by relating it to the quantum $su_{q}(1,1)$ algebra~\cite{16}. It has been found~\cite{16} that the periodicity of the time evolution of the physical quantities, which is a typical feature of the JC model with intensity-dependent coupling, is destroyed when the atom interacts with the $q$ analogue of the $SU(1,1)$ coherent state. In order to extract possible information about the physical meaning of the \textit{q}-deformation, the atomic inversion of the standard JCM with a \textit{q}-deformed field initially prepared in a \textit{maths}-type \textit{q}-deformed coherent state~\cite{53} has been studied~\cite{19}. Various versions of the JC model and their \textit{q-}deformed extensions have been treated in a unified formalism based on the generalized deformed oscillator algebra~\cite{21}. The temporal evolution of atomic inversion and quantum fluctuations of atomic dipole variables in three variants of the two-photon non-dissipative \textit{q}-deformed JC model have been studied~\cite{38} for both on-and off-resonant atom-field interaction. Particularly, it has been found that for nonzero detuning the atomic inversion exhibits superstructures, which are absent in the non-deformed JC model, and the magnitude of dipole squeezing may be increased.  By solving a dynamical problem characterized by a quite general \textit{f}-deformed JC model, it has also been proposed~\cite{41} a theoretical scheme to show the possibility of generating various families of nonlinear (\textit{f}-deformed) coherent states~\cite{54} of the radiation field in a lossless coherently pumped micromaser. In Ref.~\cite{42}, the influence of nonlinear quantum dissipation on the dynamical properties of the \textit{f}-deformed JC model in the dispersive approximation has been explored. In Ref.~\cite{44} the author has studied the influence of the intrinsic decoherence on quantum statistical properties of a nondegenerate two-photon \textit{f}-deformed JC model. Furthermore, by introducing some classes of two-parameter (\textit{p,q}) deformation of the JC model in the rotating wave approximation (RWA) the associated nonlinear vector coherent states have been constructed~\cite{46}.

Another significant and noteworthy extension of the JC model is the inclusion of atomic motion as well as the effect of the field structure of the cavity mode sustained in the cavity (see, e.g. Refs.~\cite{43}). The motivation for such studies is the recent development in the technologies of laser cooling and atom trapping to achieve cooled atoms and super-cooled atoms~\cite{55}. On the other hand, when the atomic motion is taken into account, some interesting nonlinear transient effects that are similar to self-induced transparency and adiabatic following can be observed in the atomic population~\cite{39}. Furthermore, the influences of the atomic motion and the field-mode structure on the squeezing of the field~\cite{20}, on the cavity-field entropy and Schr\"{o}dinger-cat states generation in the JC model~\cite{32}, and on the Rabi oscillations of an atom irradiated by a laser in resonance with the atomic transition~\cite{43} have been studied. In Ref.~\cite{30} the authors have provided a dressed formalism for the dynamics of a moving two-level atom interacting with a quantized cavity-field by taking into account the  Roentgen interaction~\cite{56}. All of the foregoing studies devoted to the JC model with atomic motion have been done only under the condition that the influence of the gravitational field is not taken into account. However, it has often been argued that the achievement of the cooled atoms would be a formidable task as the atoms move so slowly (with a velocity of a few millimeters or centimeters per second for a time period of several milliseconds or more) that they start exactly to be extremely sensitive to earth gravity~\cite{57}. For this reason, it is of interest to study the temporal evolution of a moving atom simultaneously exposed to the gravitational field and a single-mode cavity-field. By referring to the equivalence principle, one can get a clear picture of what is going to happen in the interacting atom-field system exposed to a classical homogeneous gravitational field~\cite{58}. Gravity action on particles of small velocities has been observed in matter wave experiments~\cite{59}. Recently, it has been identified~\cite{60} the lowest stationary quantum state of neutrons in the earth's gravitational field by measuring the neutron transmission between a horizontal mirror on the bottom and an absorber or scatterer on top. A semi-classical description of a two-level atom interacting with a running laser wave in a gravitational field has been studied in Refs.~\cite{61,62}. However, the semi-classical treatment prevent us from studying the pure quantum effects occurring in the course of atom-field interaction. In a series of recently published papers~\cite{58, 63, 64, 65} various aspects of the influence of a classical homogeneous gravitational field on quantum dynamics of the JC model have been analyzed. In Ref.~\cite{63}, within a quantum treatment of the internal and external dynamics of the atom, a theoretical scheme has been presented based on an $su(2)$ dynamical algebraic structure to investigate the influence of a classical homogeneous gravitational field on the quantum non-demolition measurement of atomic momentum in the dispersive JC model. The effects of the gravitational field on quantum statistical properties of the lossless as well as the phase-damped JC models have been studied in Refs.~\cite{58} and ~\cite{64}, respectively. Also, the effects of gravity on the dynamical evolution of the cavity-field entropy and the creation of the Schr\"{o}dinger-cat state in the Jaynes-Cummings model have been examined in Ref.~\cite{65}. The results reveal that the gravitational field seriously suppresses non-classical properties of both the cavity-field and the moving atom.

Motivated by the above-mentioned studies on various versions of generalized JC models, in the present paper we deal with the question of how does the gravitational field affect nonclassical statistical properties of the \textit{f}-deformed(nonlinear) JC model. The system under consideration consists of a moving two-level atom exposed to a classical homogeneous gravitational field and interacting with an \textit{f}-deformed single-mode quantized cavity-field including the nonlinearities of both the field and the intensity-dependent atom-field coupling. Formally, the model Hamiltonian for the atom-field system has the same structure as non-deformed JC model in the presence of gravity~\cite{58, 63, 64, 65} with the photon annihilation and creation operators   $\hat{a}$ and $\hat{a}^+$ replaced, respectively, by the \textit{f}-deformed operators $\hat{A}$ and $\hat{A}^+$ obeying the deformed commutation relation $[\hat A,\hat A^ +  ] = (\hat n + 1)f^2(\hat n + 1) - \hat n f^2(\hat n)$. The nonlinearity (deformation) $f(\hat n)$ function plays an important role in our treatment since it determines the form of nonlinearities of both the cavity-field and the intensity-dependent atom-field coupling. This paper is organized as follows. In section 2, we present a quantum treatment of the internal and external dynamics of the moving two-level atom and find an alternative \textit{f}-deformed $su(2)$ dynamical algebraic structure within the system. Based on this deformed $su(2)$ structure, we obtain an effective Hamiltonian describing the nonlinear atom-field interaction in the presence of a classical homogeneous gravitational field. In section 3, by solving the Schr\"{o}dinger equation in the interaction picture, we investigate the dynamical evolution of the system under consideration and show that how the gravitational field may affect the dynamical properties of the JC model. In section 4, we study the influence of the classical gravitational field on the quantum statistical properties of the two-level atom and the \textit{f}-deformed quantized radiation field. Considering the cavity-field to be initially in a \textit{maths}-type \textit{q}-deformed coherent state, associated with the nonlinearity function ~$f(\hat n) = \sqrt {\frac{1}{{\hat n}}\frac{{1 - q^{\hat n} }}{{1 - q}}}$,  and the two-level atom in a coherent superposition of the ground and excited states, we investigate the temporal evolution of the atomic population inversion, atomic dipole squeezing, atomic momentum diffusion, photon counting statistics, and  deformed quadrature squeezing of the radiation field. Finally, we summarize our conclusions in section 5.

\section{$f$-deformed JC model in the presence of classical gravity}

We consider the interaction of a single two-level moving atom of mass $M$ characterized by the Pauli operators $\hat\sigma_z,\,\hat\sigma_{\pm}$, the momentum and position operators $\hat{\vec p}$ and $\hat{\vec x}$  with an \textit{f}-deformed single-mode traveling field characterized by the \textit{f}-deformed annihilation and creation operators $\hat{A},\,\hat{A}^+$, respectively, in the presence of a classical gravitational field. Following Refs.~\cite{38,58}, the total Hamiltonian of the system under the RWA can be written as

\begin{equation}\label{1}
\begin{array}{rcl}
\hat H = \frac{\hat{\vec{p}}^2}{2M} - M\vec g.\hat{\vec{x}} + \frac{\hbar}{2}\omega\hat \sigma _z  + \hbar \nu \hat A^ +  \hat A + \hbar \lambda (e^{i\vec k.\hat{\vec{x}}} \hat \sigma _ +  \hat A + e^{ - i\vec k.\hat{\vec{x}}}\hat A^ +  \hat \sigma _ -  ),
\end{array}
\end{equation}
where $\omega$ denotes the atomic transition frequency, $\nu$ is the frequency of the radiation field, the atom-field coupling constant $\lambda$ is a real number, $\vec{k}$ is the wave-vector of the traveling radiation field, and $\vec g$ denotes the constant gravitational acceleration acting on the atom. The operators $\hat{A},\,\hat{A}^+$ are deformed annihilation and creation operators constructed from the usual bosonic operators  $\hat{a},\,\hat{a}^+\,([\hat{a},\hat{a}^+] =1)$ and number operator $\hat n=\hat{a}^+\hat{a}$  as follows

 \begin{equation}\label{2}
\begin{array}{rcl}
 \hat{A}=\hat{a}f(\hat{n}),
  \hat{A}^+=f(\hat{n})\hat{a}^+,
\end{array}
\end{equation}
in which the deformation function $f(\hat{n})$ is an arbitrary real function of $\hat n$. The deformed operators $\hat{A},\,\hat{A}^+$ satisfy the deformed bosonic oscillator commutation relations
\begin{equation}\label{3}
\begin{array}{rcl}
[\hat A,\hat A^ +  ] = (\hat n + 1)f^2(\hat n + 1) - \hat n f^2(\hat n),~ [\hat A,\hat n] = \hat A,~ [\hat A^ +  ,\hat n] =  - \hat A^ + .
\end{array}
\end{equation}
The knowledge of the function $f(\hat{n})$ determines all the properties of the deformed algebra (\eqn{3}). In the limiting case $f(\hat{n})=1$, the Hamiltonian (\eqn{1}) reduces to the non-deformed JC model in the presence of atomic motion and gravity~\cite{38} and the algebra (\eqn{3}) reduces to the well-known Heisenberg-Weyl algebra generated by $\hat{a},\,\hat{a}^+$ and the identity $\hat I$:

\begin{equation}\label{4}
\begin{array}{rcl}
[\hat a,\hat a^ +  ] = \hat I,~ [\hat a,\hat n] = \hat a,~ [\hat a^ +  ,\hat n] =  - \hat a^ + .
\end{array}
\end{equation}

To achieve a more clear insight to the physical meaning of the \textit{f}-deformed JC Hamiltonian given in Eq.(\eqn{1}) it is proper to rewrite it in terms of  non-deformed field operators $\hat{a},\,\hat{a}^+$. Using (\eqn{2}) we arrive at
\begin{equation}\label{5}
\begin{array}{rcl}
\hat H = \frac{\hat{\vec{p}}^2}{2M} - M\vec g.\hat{\vec{x}}+
+ \frac{\hbar}{2}\omega \hat \sigma _z  +\hbar\nu \hat a^ +  \hat a +\hbar \nu (f^2 (\hat n)-1)\hat a^ +  \hat a
+ \hbar \lambda (e^{i\vec k.\hat{\vec{x}}} \hat \sigma _ +  \hat af(\hat n)
+ e^{ - i\vec k.\hat{\vec{x}}} f(\hat n)\hat a^ +  \hat \sigma _ -  ).
\end{array}
\end{equation}
In this way, we learn that the above Hamiltonian describes an intensity-dependent interaction of a single two-level moving atom, exposed to a classical homogeneous gravitational field,
with a non-deformed single-mode radiation field in the presence of an additional nonlinear interaction represented by the fifth term in Eq.(\eqn{5}).
As a well-known example if we choose $ f(\hat a^+\hat a)=\sqrt{1+k(\hat a^+\hat a-1)}$
where $k$ is a positive constant, the fifth term in Eq.(\eqn{5}) takes the form $\chi\hat{a}^{+2} \hat{a}^2 $
which is reminiscent of the Kerr-induced interaction \cite{66} with $\chi=\hbar k\nu$  as the dispersive part of the third-order nonlinearity of
the Kerr-like medium. Therefore, in this case the model consists of a single two-level moving atom exposed simultaneously to a classical
homogeneous gravitational field and a single-mode field surrounded by a nonlinear Kerr-like medium contained inside a lossless cavity.
Furthermore, it is easily seen that with the above choice for the function $f(\hat{n})$
the commutation relations (\eqn{3}) take the form $[\hat A,\hat A^ +  ] = 2\hat A_0,\, [\hat A_0 ,\hat A^ +  ] = k\hat A^ +,\,[\hat A_0 ,\hat A] =  - k\hat A$, with $\hat A_0  = \frac{1}{2} + k\hat a^ +  \hat a$. It is noteworthy that these relations define $su(1,1)$ algebra when $ k =1$.

It is apparent that there exist a constant of motion in Hamiltonian (\eqn{5}),
\begin{equation}\label{6}
\begin{array}{rcl}
\hat c = \hat n + \left| e \right\rangle \left\langle e \right|,
\end{array}
\end{equation}
that is
\begin{equation}\label{7}
\begin{array}{rcl}
[\hat c,\hat H] = [\hat c,\hat n] = [\hat c,\hat\sigma _z ] = 0.
\end{array}
\end{equation}
By using the constant of motion $\hat c$, which represents the total number of excitations of the atom-radiation system, one can expand any function   $U(\hat n)$ as follows
\begin{eqnarray}\label{8}
 U(\hat n) &=& U(\hat c - \left| e \right\rangle \left\langle e \right|) \nonumber\\
 &=& \sum\limits_{i = 0} {\frac{{U^{(i)} (0)}}{{i!}}} (\hat c - \left| e \right\rangle \left\langle e \right|)^i\nonumber\\
  &=& \sum\limits_{i = 0} {\frac{{U^{(i)} (0)}}{{i!}}} (\hat c^i  - i\hat c^{i - 1} \left| e \right\rangle \left\langle e \right| + \frac{{i(i + 1)}}{{2!}}\hat c^{i - 2} (\left| e \right\rangle \left\langle e \right|)^2  + ..)\\
  &=& \sum\limits_{i = 0} {\frac{{U^{(i)} (0)}}{{i!}}} (\hat c^i  + \{ (\hat c - 1)^i  - \hat c^i \} \left| e \right\rangle \left\langle e \right|) \nonumber \\
   &=& U(\hat c) + \{ U(\hat c - 1) - U(\hat c)\} \left| e \right\rangle \left\langle e \right|.\nonumber
\end{eqnarray}
By considering~$U(\hat n) = \hat nf^2 (\hat n)$ we obtain
\begin{eqnarray}\label{9}
\hat nf^2 (\hat n) &=& (\hat c - \left| e \right\rangle \left\langle e \right|)(f^2(\hat c)  + \{ f^2 (\hat c - 1) - f^2 (\hat c)\} \left| e \right\rangle \left\langle e \right|)\nonumber \\
&=&\frac{1}{2}\{\hat cf^2 (\hat c) + (\hat c - 1)f^2 (\hat c - 1)\} + \frac{1}{2}\{ (\hat c - 1)f^2 (\hat c - 1) - \hat cf^2 (\hat c)\} \hat \sigma _z .
\end{eqnarray}
By making use of Eq. (\eqn{9}) in the Hamiltonian (\eqn{5}) and after some rearrangement, we arrive at
\begin{equation}\label{10}
\begin{array}{rcl}
\hat H = \frac{\hat{\vec{p}}^2}{2M} - M\vec g.\hat{\vec{x}}
+ \frac{\hbar}{2}\hat\Delta (\hat c)\hat \sigma _z
+\hbar \lambda (e^{i\vec k.\hat{\vec{x}}} \hat \sigma _
+  \hat af(\hat n) + e^{ - i\vec k.\hat{\vec{x}}} f(\hat n)\hat a^ +  \hat \sigma _ -  )+\hat H_c,
\end{array}
\end{equation}
where
\begin{equation}\label{11}
\begin{array}{rcl}
\hat H_c  = \frac{\hbar}{2}\nu\{\hat c f^2(\hat c)  + (\hat c - 1)f^2(\hat c - 1) \},
\end{array}
\end{equation}
and
\begin{equation}\label{12}
\begin{array}{rcl}
\hat\Delta (\hat c) = \omega  - \nu \{ \hat c f^2(\hat c)  - (\hat c - 1)f^2(\hat c - 1) \}.
\end{array}
\end{equation}
We now introduce the following operators
\begin{equation}\label{13}
\begin{array}{rcl}
\hat S_ + ^d  = \hat \sigma ^ +  \hat af(\hat n)\frac{1}{\sqrt{\hat{c}}},~\hat S_ - ^d  =\frac{1}{\sqrt{\hat{c}}} f(\hat n)\hat a^ +  \hat \sigma ^ -  ,~\hat S_0  =\frac{1}{2}\{ \left| e \right\rangle \left\langle e \right| - \left| g \right\rangle \left\langle g \right|\}  = \frac{1}{2}\hat \sigma _z ,
\end{array}
\end{equation}
which satisfy the following commutation relations
\begin{equation}\label{14}
\begin{array}{rcl}
[\hat S_ - ^d ,\hat S_ +  ^d ] =  - 2\xi (\hat n;\left| e \right\rangle \left\langle e \right|;\left| g \right\rangle \left\langle g \right|)\hat S_0 ,~[\hat S_0 ,\hat S_ \pm ^d ] =  \pm \hat S_ \pm ^d ,
\end{array}
\end{equation}
where
\begin{equation}\label{15}
\begin{array}{rcl}
\hat\xi (\hat n;\left| e \right\rangle \left\langle e \right|;\left| g \right\rangle \left\langle g \right|) =\frac{1}{\hat{c}}\{ f^2 (\hat n+1)(\hat n + 1)\left| e \right\rangle \left\langle e \right| + f^2 (\hat n)\hat n\left| g \right\rangle \left\langle g \right|\}.
\end{array}
\end{equation}
The operators~$\hat S_ \pm ^d,~\hat S_0 $ are the generators of an \textit{f}-deformed~$su(2)$ algebra. In the terms of these generators, the
Hamiltonian (\eqn{10}) can be rewritten as
\begin{equation}\label{16}
\begin{array}{rcl}
\hat H = \frac{\hat{\vec{p}}^2}{2M} - M\vec g.\hat{\vec{x}} + \hbar\hat\Delta (\hat c)\hat S _0
+\hbar \lambda (e^{i\vec k.\hat{\vec{x}}} \hat S_ +  ^d \sqrt{\hat{c}}+ \sqrt{\hat{c}}\;e^{ - i\vec k.\hat{\vec{x}}} \hat S_ - ^d  )+\hat H_c.
\end{array}
\end{equation}
The dynamical evolution of the system can be investigated by the Hamiltonian (\eqn{16}). For solving Eq.(\eqn{16}) we introduce the time-dependent unitary operator $\hat T=\exp (i M\vec g.\hat{\vec x} t/\hbar )\exp ( - i\vec k.\hat{\vec x}\hat S_0 )$ and we have
\begin{equation}\label{17}
\begin{array}{rcl}
\hat H_e  = \hat T \hat H \hat T^{-1}= \hat H_0  + \hat H_1,
\end{array}
\end{equation}
where
\begin{equation}\label{19}
\begin{array}{rcl}
\hat H_0  = \frac{{(\hat {\vec p} + M\vec g t)^2 }}{{2M}} + \hbar \hat \Delta_{1} (\hat c,t)\hat S_0  + \hat H_c  + \frac{{\hbar ^2 \vec{k}^2 }}{{8M}},
\end{array}
\end{equation}
and
\begin{equation}\label{20}
\begin{array}{rcl}

\hat H_1  = \hbar \lambda (\sqrt{\hat{c}}\hat  S_ - ^d    + \hat S_ +^d \sqrt{\hat{c}} ).

\end{array}
\end{equation}
Here we have introduced the operator $\hat \Delta_{1} (\hat c,t)$ as the \textit{f}-deformed Doppler shift detuning at time \textit{t},

\begin{equation}\label{21}
\begin{array}{rcl}
\hat \Delta_{1} (\hat c,t) = \hat\Delta (\hat c) + \frac{\vec k.\hat{\vec p}}{M} + \vec k.\vec g t.
\end{array}
\end{equation}
Thus, due to the presence of the gravitational field~($\vec k.\vec g t$), the nonlinear associated with \textit{f}-deformation~($f(\hat n)$), and the Doppler shift~($\vec k.\hat{\vec p}/M $) the detuning between the cavity-field and the atomic transition frequency is modified. Finally, in the interaction picture the transformed Hamiltonian (\eqn{20}) takes
the following form

\begin{equation}\label{22}
\begin{array}{rcl}
\hat H_{{\mathop{\rm int}} }  = e^{\frac{i}{\hbar}\int\hat H_0dt } \hat H_1 e^{ - \frac{i}{\hbar}\int\hat H_0dt}=
 \hbar \lambda (\sqrt{\hat{c}}\;e^{ - i\hat \Delta (\hat c,t)t} \hat S_ - ^d  + e^{i\hat \Delta (\hat c,t)t} \hat S_ + ^d \sqrt{\hat{c}}),

\end{array}
\end{equation}
where $\hat \Delta (\hat c,t) = \hat\Delta (\hat c) + \vec k.\hat{\vec p}/M + \vec k.\vec g t/2$. By using the above Hamiltonian we can investigate the dynamical evolution of the system under consideration.
This will be done in the two following sections.
\section{Dynamical evolution of the system}

In the previous section, we obtained an effective Hamiltonian for the interaction of a two-level atom with an \textit{f}-deformed quantized radiation field, in the presence of a classical homogenous gravitational field. In the present section, we investigate the temporal evolution of the system. For this purpose, we proceed to solve the Schr\"{o}dinger equation for state vector~$|\psi(t)\rangle$ ; i.e.
\begin{equation}\label{24}
\begin{array}{rcl}

i\hbar \frac{{\partial \left| {\psi (t)} \right\rangle }}{{\partial t}} = \hat H_{{\mathop{\rm int}} } \left| {\psi (t)} \right\rangle.
\end{array}
\end{equation}
At any time $t$, the state vector~$|\psi(t)\rangle$  is a linear combination of the states~$|e,n\rangle\bigotimes|\vec p\rangle$ and~$|g,n\rangle\bigotimes|\vec p\rangle$. Here~$|e,n\rangle$($|g,n\rangle$) is the state in which the atom is in the excited state~$|e\rangle$(the ground state $|g\rangle$) and the cavity-field has~$n$ photons. Now we can propose the following ansatz for~$|\psi(t)\rangle$
\begin{equation}\label{25}
\begin{array}{rcl}
\left| {\psi (t)} \right\rangle  = \int {d^3 p} \sum\limits_{n }^{} {[\psi _{1,n} (\vec p,t)\left| {e,n} \right\rangle  \otimes \left| {\vec p} \right\rangle  + \psi _{2,n+1
} (\vec p,t)\left| {g,n+1} \right\rangle  \otimes \left| {\vec p} \right\rangle ]}.

\end{array}
\end{equation}
Since the interaction Hamiltonian $\hat H_{int}$ couples only the states~$|e,n\rangle\otimes|\vec p\rangle $ and~$|g,n+1\rangle\otimes|\vec p\rangle$, we consider the evolution of the amplitudes~$\psi_{1,n},~\psi_{2,n+1}$. By substituting equation (\eqn{25}) into (\eqn{24}),
the following equations of motion for the time-dependent probability amplitudes~$\psi_{1,n},~\psi_{2,n+1}$ are obtained

\begin{eqnarray}\label{26}
\frac{d}{d t}\psi _{1,n} &=&  - i\lambda \sqrt {n + 1} f(n + 1)e^{i\Delta (n,t)t} \psi _{2,n+1} ,\nonumber\\
\frac{d}{d t}\psi _{2,n+1} &=&  - i\lambda \sqrt {n + 1} f(n + 1)e^{ - i\Delta (n,t)t} \psi _{1,n} ,
\end{eqnarray}
where, we have introduced the modified detuning parameter $\Delta (n,t)$ as
\begin{equation}\label{27}
\begin{array}{rcl}
\Delta (n,t) = \Delta (n) + \frac{\vec k.\vec p}{M} + \frac{\vec k.\vec g t}{2},
\end{array}
\end{equation}
with
\begin{equation}\label{28}
\begin{array}{rcl}
\Delta (n) = \Delta  + \nu \{ nf^2 (n) - (n + 1)f^2 (n + 1) + 1\},
\end{array}
\end{equation}
and

\begin{equation}\label{29}
\begin{array}{rcl}
\Delta=\omega- \nu,
\end{array}
\end{equation}
is the usual detuning parameter.\\
The two coupled first-order differential equations (\eqn{26}) can be solved in a straightforward way, so we have

\begin{eqnarray}\label{30}
\psi _{2,n+1} (t) &=& e^{ - it(\Delta (n,t))} \{ c_{1,n} H( - 1 - 2ia^2 (n),b(t)) + c_{2,n} ~{}_1F_1 (1/2 + ia^2 (n),\frac{1}{2},b^2 (t))\},\nonumber\\
\psi _{1,n} (t) &=& c_{1,n} H( - 2ia^2 (n),b(t)) + c_{2,n} ~{}_1F_1 (ia^2 (n),\frac{1}{2},b^2 (t)),
\end{eqnarray}
where~$H$ and ${}_1F_1$ denote, the Hermit and the confluent hyper-geometric functions, respectively, and by definition
\begin{equation}\label{31}
\begin{array}{rcl}
a^2 (n) = \frac{{\Omega ^2 (\hat p,n,g) - \Delta ^2 _k }}{{\vec k.\vec g}} - i,b(t) = \frac{{( - 1)^{1/4} (\vec k.\vec gt + \Delta _k )}}{{2\sqrt {\frac{\vec k.\vec g}{2}} }},
\Delta _k  = \Delta (n) +\frac{\vec k.\vec p}{M},

\end{array}
\end{equation}
in which
\begin{equation}\label{32}
\begin{array}{rcl}
\Omega ^2 (\vec p,n,\vec g) = \Omega ^2 (\vec p,n,0) + i\vec k.\vec g,\\
\end{array}
\end{equation}
with $\Omega ^2 (\vec p,n,0) = \lambda ^2 (n + 1)f^2 (n + 1) + \Delta ^2 _k$, is the \textit{f}-deformed gravity-dependent Rabi frequency.

At time~$t = 0$ the atom is uncorrelated with the field and the state vector of the system can be
written as a direct product
\begin{equation}\label{33}
\begin{array}{rcl}
\left| {\psi (0)} \right\rangle  = \left| {\psi (0)} \right\rangle _{atom}  \otimes \left| {\psi (0)} \right\rangle _{field}  \otimes \left| {\psi (0)} \right\rangle _{c.m}  = \int {d^3 p\phi (\vec p)} \left| {\vec p} \right\rangle  \otimes \sum\limits_{n = 0} {(w_n c_e \left| {e,n} \right\rangle  + w_n c_g \left| {g,n} \right\rangle )},

\end{array}
\end{equation}
where we have assumed that initially the cavity-field is in a coherent superposition of Fock states ($\sum\limits_n {w(n)\left| n \right\rangle}$), the two-level atom is in a coherent superposition of its excited and ground states, and the state vector for the center-of-mass degree of freedom is~$\left| {\psi (0)} \right\rangle _{c.m}  = \int {d^3 p\phi (\vec p)} \left| {\vec p} \right\rangle$.\\
Thus, the time-independent coefficients $c_{1,n}=c(1)/c$ and $c_{2,n}=c(2)/c$ in Eqs.(\eqn{30}) are obtained as

\begin{eqnarray}\label{36}
c &=& {}_1F_1 (\frac{1}{2} + ia^2 (n),\frac{1}{2},b^2 (0))H( - 2ia^2 (n),b(0)) - {}_1F_1 (ia^2 (n),\frac{1}{2},b^2 (0))H( - 1 - 2ia^2 (n),b(0)),\nonumber\\
c(2) &=&  H( - 1 - 2ia^2 (n),b(0))\psi _1 (0) - H( - 2ia^2 (n),b(0))\psi _2 (0),\\
c(1)&=&  {}_1F_1 (\frac{1}{2} + ia^2 (n),\frac{1}{2},b^2 (0))\psi _1 (0) - {}_1F_1 (ia^2 (n),\frac{1}{2},b^2 (0))\psi _2 (0)\nonumber.
\end{eqnarray}
In the next section, by using the sate vector (\eqn{25}), we examine the temporal evolution of quantum statistical properties of the \textit{f}-deformed interacting atom-field system in the presence of a classical homogeneous gravitational field.
\section{quantum statistical properties of the system}
In this section we shall study the influence of the classical gravity field on the dynamical evolution of quantum statistical properties of the \textit{f}-deformed JC model, including the atomic population inversion, atomic dipole squeezing, atomic momentum diffusion, photon-counting statistics, and deformed quadrature squeezing of the field.
\subsection{Atomic population inversion}
Here, we discuss the collapse-revival phenomenon in the evolution of the atomic inversion for the state vector $ |\psi(t)\rangle$. This phenomenon is a purely quantum mechanical effect and has its origin in the granular structure of the photon number distribution of the initial field~\cite{2}. The phenomenon has been realized experimentally in the sense that the state of the atomic beam leaving the cavity is monitored by ionization detectors~\cite{68}. The revival-collapse phenomenon has also been seen in nonlinear optics for the single-mode mean photon number of the Kerr nonlinear coupler~\cite{69} when the modes are initially prepared in coherent state. At present this phenomenon is actively discussed in quantum optics not only for atomic system applications but also for the cases of low-dimension quantum nanostructures~\cite{70}, the Landau-Zener process ~\cite{71}, and trapped atoms~\cite{72}. For the model under consideration the atomic population inversion at time $t$ is obtained as follows

\begin{equation}\label{37}
\begin{array}{rcl}
W(t) = \left\langle {\psi (t)} \right|\hat{\sigma} _z \left| {\psi (t)} \right\rangle=
\int {d^3 p} \sum\limits_{n = 0}^\infty  {\left\{ {|\psi _{1,n} |^2  - |\psi _{2,n} |^2 } \right\}} .
\end{array}
\end{equation}
We assume at $t = 0$, the two-level atom is in a coherent superposition of the excited state and the ground state with~$c_e(0)=\frac{1}{\sqrt{2}},~c_g(0)=\frac{1}{\sqrt{2}}$
and the cavity-field is prepared in a \textit{maths}-type \textit{q}-deformed coherent state, $\left| {\psi (0)} \right\rangle _f  = \left| {\alpha ,f} \right\rangle  = \sum\limits_{n = 0}^\infty  {w(n)} \left| n \right\rangle$ in which $
w(n) = N_\alpha  \sum\limits_{n = 0}^\infty  {\frac{{\alpha ^n }}{{\sqrt {[nf^2 (n)]!} }}}$ with $
N_\alpha   = \left( {\sum\limits_{n = 0}^\infty  {\frac{{|\alpha |^{2n} }}{{[nf^2 (n)]!}}} } \right)^{ - 1/2}$ and ~$f(\hat n) = \sqrt {\frac{1}{{\hat n}}\frac{{1 - q^{\hat n} }}{{1 - q}}}$. This type of deformed coherent states can be identified as a family of nonlinear  coherent states which can be generated in a lossless coherently pumped micromaser with an intensity-dependent atom-field coupling governed by the nonlinearity function $f(n) = \sqrt {\frac{1}{{n}}\frac{{1 - q^{n} }}{{1 - q}}}$ \cite{41}. From the point of view of the photon statistics, this family of states exhibits some nonclassical features such as antibunching and quadrature squeezing for certain ranges of the deformation  parameter \textit{q} \cite{38,41}.\\
In Figs.1(a)-1(c) we have plotted the behaviour of $W(t)$ as a function of the scaled time $\lambda t$ for three different values of  the parameter $\vec k.\vec g$ and for $q=1.04$. In this figure and all the subsequent figures we set $M = 10^{ - 26} kg,\,\Delta _k  = 3 \times 10^7 rad/s,\,\alpha  = 2,\,\lambda  = 10^5 rad/s,\,k = 10^7 m^{ - 1}, $ and $\phi (\vec p) = \frac{1}{{\sqrt {2\pi } }}\exp ( - p^2 )$ ~\cite{61,62}. Here, it is necessary to point out that the relevant time scale introduced by the gravitational influence is $\tau_a=1/\sqrt{\vec k.\vec g}$ ~\cite{63}. Thus for an optical field with $|\vec k|=10^7 m^{-1}$ we have $\tau_a\cong10^{-4} s$ . It is important to note that the single deformation parameter \textit{q} determines the structure of initial deformed coherent state of the cavity-field. Furthermore, it may be viewed as a phenomenological constant controlling the strength of intensity-dependent atom-field coupling as well as the interaction between the field and nonlinear medium contained inside the cavity. As an interesting point we note that when~$q = 1 \pm \varepsilon (0 < \varepsilon  <  < 1)$, the nonlinearity function $f(\hat n)$ reduces to $f(\hat n) \simeq\sqrt {1 \pm \frac{\varepsilon}{2} (\hat n - 1)}$. Therefore one can infer that up to the first order approximation the nonlinearity of the model under consideration may be described as a Kerr-type nonlinearity. Figure.1(a) displays the case when the gravitational influence is negligible~($\vec k.\vec g=0$). This means very small~$\vec k.\vec g$ , i.e., the momentum transfer from the radiation field to the atom is only slightly altered by the gravitational acceleration because the latter is very small or nearly perpendicular to the wave-vector $\vec k$.

Figures.1(b) and 1(c) illustrate the case when we consider the gravitational influence for~$\vec k.\vec g=2\times10^7 sec^{-2}$ and $\vec k.\vec g=8\times10^7 sec^{-2}$, respectively. As it is seen from Fig.1(a), in the absence of gravitational influence the Rabi-like oscillations can be identified. Furthermore, the atomic inversion oscillates around a positive nonzero value which means that due to the deformation more energy is stored in the atomic subsystem. Physically, it is due to the change in energy-level structure of the deformed model under consideration. Increasing of the gravitational influence (Figs.1(b), 1(c)) induces the population inversion to oscillate so drastically that the phenomenon of collapse-revival is not so clear. The emergence of this quasi-chaotic behaviour of atomic population inversion can be attributed to a detuning that varies linearly with time [Eq.(\eqn{27})] and a modified gravity-dependent Rabi frequency [Eq.(\eqn{32}]. It has been shown \cite{58} that these continuous changes in oscillation frequencies influence the collapse and revival times such that subsequent revivals overlap and ultimately a chaotic-like behaviour of time evolution of population inversion occurs. In addition, with the increasing value of the parameter~$\vec k.\vec g$ the atomic inversion oscillates around a negative nonzero value, i.e., due to the gravitational field more energy is stored in the cavity-field. Besides, numerical analysis shows that by increasing the deformation (nonlinearity) parameter \textit{q}, the population inversion shows chaotic-like behaviour, for both cases of $\vec k.\vec g=0$ and $\vec k.\vec g\neq0$. The origin of this chaotic-like behaviour can also be understood by examining the effect of deformation on the collapse and revival times \cite{38}.

\subsection{Atomic dipole squeezing}
The squeezing of quantum fluctuations is one of the most fundamental manifestations of the Heisenberg uncertainty relations, which are among the most important principles of quantum mechanics. Squeezing, in particular that of the radiation field, has attracted a great deal of interest and been extensively studied both theoretically and experimentally over the past thirty years due to its practical applications in many optical devices as well as in the quantum information, e.g. quantum teleportation \cite{73}, power-recycled interferometer \cite{76} and phase-modulated signal recycled interferometer \cite{77}. Meanwhile, increased attention has also been devoted to the squeezing of the fluctuations of the atomic dipole variables \cite{78} owing to its potential applications in the high-resolution spectroscopy \cite{79}, the high-precision atomic fountain clock \cite{80}, the high-precision spin polarization measurements \cite{81}, etc. Moreover, the important connection of atomic dipole squeezing with the squeezing of the radiation field has been established \cite{82}; atomic dipole squeezing is shown to produce reduced quantum fluctuations in the radiated field.
	In order to analyze the atomic dipole squeezing in the model under consideration, we consider the two slowly varying Hermitian quadrature operators

\begin{equation}\label{40}
\begin{array}{rcl}
\hat \sigma _x (t) = \frac{1}{2}(\hat \sigma ^ +  e^{ - i\omega t}  + \hat \sigma ^ -  e^{i\omega t} ),\,\,\hat \sigma _y (t) = \frac{1}{{2i}}(\hat \sigma ^ +  e^{ - i\omega t}  - \hat \sigma ^ -  e^{i\omega t} ),
\end{array}
\end{equation}
In fact~$\hat \sigma _x$, and $\hat \sigma _y $ correspond to the dispersive and absorptive components of the amplitude of the atomic polarization, respectively. They obey the commutation relation~$[\hat \sigma _x ,\hat \sigma _y ] = \frac{i}{2}\hat \sigma _z$. Correspondingly, the Heisenberg uncertainty relation is $(\Delta \hat \sigma _x )^2 (\Delta \hat \sigma _y )^2  \ge \frac{1}{{16}}\left| {\left\langle {\hat \sigma _z } \right\rangle } \right|^2$, where $(\Delta \hat \sigma _i )^2  = \left\langle {\hat \sigma _i^2 } \right\rangle  - \left\langle {\hat \sigma _i } \right\rangle ^2$ is the variance in the component~$\hat \sigma _i (i = x,y)$ of the atomic dipole. The fluctuations in the component~$\hat \sigma _i (i = x,y)$ are said to be squeezed \cite{hil} if the variance in $\hat \sigma _i$ satisfies the condition $(\Delta \hat \sigma _i )^2  < \frac{1}{4}\left| {\left\langle {\hat \sigma _z } \right\rangle } \right|{\rm  , }(i = x\,\,or\,\, y)
$. Since $\left\langle {\hat \sigma _i^2 } \right\rangle  = 1/4$ this condition may be written as
\begin{equation}\label{41}
\begin{array}{rcl}
F_i (t) = 1 - 4\left\langle {\hat \sigma _i (t)} \right\rangle ^2  - \left| {\left\langle {\hat \sigma _z (t)} \right\rangle } \right| < 0,{\rm    }(i = x\,\,or\,\,y)
\end{array}
\end{equation}
The expectation values of the atomic operators $\hat{\sigma}_x$ and $\hat{\sigma}_y$ in the state $|\psi(t)\rangle$ of the atom-field system, given by (\eqn{25}), are
\begin{equation}\label{42}
\begin{array}{rcl}
\left\langle {\hat \sigma _x (t)} \right\rangle  = \int {d^3 } p\sum\limits_{n = 0}^\infty  {{\mathop{\rm Re}\nolimits} \left( {\psi _{1,n} (t)\psi _{2,n}^ *  (t)\exp ( - i\omega t} \right)},
\end{array}
\end{equation}
\begin{equation}\label{42.5}
\begin{array}{rcl}
\left\langle {\hat \sigma _y (t)} \right\rangle  = \int {d^3 } p\sum\limits_{n = 0}^\infty  {{\mathop{\rm Im}\nolimits} \left( {\psi _{1,n} (t)\psi _{2,n}^ *  (t)\exp ( - i\omega t} \right)}.
\end{array}
\end{equation}
Let us examine numerically the squeezing properties of the atomic dipole variable $\hat{\sigma}_y$ in the model under consideration. Figures. 2(a)-2(c) show the time evolution of $F_y(t)$, corresponding to the squeezing of the component $\hat{\sigma}_y$, as a function of the scaled time $\lambda t$, when we consider the same values of parameters taken in Figs.1. As it is seen, when the influence of the gravitational field increases, the dipole squeezing is weakened. Also, numerical analysis reveals that increasing of the nonlinearity (increasing the value of the deformation parameter \textit{q}) seriously reduces the dipole squeezing.

\subsection{Atomic momentum diffusion}
The next quantity we consider is the atomic momentum diffusion. As a consequence of the atomic momentum diffusion, the atom experiences light-induced forces (radiation force) during its interaction with the radiation field. The atomic momentum diffusion is given by
\begin{equation}\label{43}
\begin{array}{rcl}

\Delta p(t) = (\left\langle {p(t)^2 } \right\rangle  - \left\langle {p(t)} \right\rangle ^2 )^{1/2}.

\end{array}
\end{equation}
By using equation(\eqn{25}), we obtain
\begin{equation}\label{43}
\begin{array}{rcl}
\Delta p(t) = \{ \sum\limits_{n = 0} {\int {d^3 pp^2 } } (\left| {\psi _{1,n} } \right|^2  + \left| {\psi _{2,n} } \right|^2 ) - [\sum\limits_{n = 0} {\int {d^3 pp} } (\left| {\psi _{1,n} } \right|^2  + \left| {\psi _{2,n} } \right|^2 )]^2 \} ^{1/2}.
\end{array}
\end{equation}
Now we turn our attention to examine numerically the dynamics of the atomic momentum diffusion. The numerical results of the evolution of $\Delta p(t)$ are shown in Figs. 3(a)-3(c) with the same corresponding data used in Fig.1.\\
Figure 3(a) illustrates the case when the gravitational influence is negligible, while Figs. 3(b) and 3(c) display the cases when we consider the gravitational influence. It is seen from Fig. 3(a) that, when the gravitational influence is negligible, the Rabi-like oscillations can be identified. However, by increasing the gravitational influence (Figs. 3(b) and 3(c)) the momentum diffusion oscillates so drastically that the phenomenon of collapse-revival is not so clear. Moreover, the atom can experience larger light-induced forces during its interaction with the radiation field, when the gravitational field increases. Further numerical analysis shows that, for both cases of $\vec k.\vec g=0$ and $\vec k.\vec g\neq0$, with the increasing value of the deformation parameter \textit{q} the momentum diffusion increases and it exhibits a chaotic-like behaviour.

\subsection{Photon-counting statistics}
One of the best-known nonclassical effects of the light field is sub-Poissonian photon statistics. The experiments on sub-Poissonian statistics are concerned with the intensity or photon-number fluctuations of the electromagnetic field. To understand the influence of gravity on sub-Poissonian statistics of the cavity-field in the \textit{f}-deformed JC model, we examine the behaviour of the second-order correlation function for the cavity-field defined by \cite{68}
\begin{equation}\label{44}
\begin{array}{rcl}
G^{(2)}(t)\equiv\frac{\langle \hat n^2(t)\rangle-\langle \hat n(t)\rangle}{(\langle \hat n(t)\rangle)^2}
 = \frac{{\sum\limits_{n = 0}^\infty  {n^2 p(n,t)}  - \sum\limits_{n = 0}^\infty  {np(n,t)} }}{{[\sum\limits_{n = 0}^\infty  {np(n,t)]^2 } }},
\end{array}
\end{equation}
where the photon probability distribution function at time $t$, ~$p(n,t)$, is given by
\begin{equation}\label{45}
\begin{array}{rcl}
p(n,t) = |\left\langle {n}
 \mathrel{\left | {\vphantom {n {\psi (t)}}}
 \right. \kern-\nulldelimiterspace}
 {{\psi (t)}} \right\rangle |^2  = \int {d^3 p\left( {|\psi _{1,n} (t)|^2  + |\psi _{2,n} (t)|^2 } \right)}.
\end{array}
\end{equation}
The photon statistics is said to be sub-Poissonian if~$G^{(2)}(t)<1$ and super-Poissonian if~$G^{(2)}(t)> 1$. The value~$G^{(2)}(t)=1$ stands for Poissonian statistics.

In Figs. 4(a)-4(c) we have sketched the temporal evolution of~$G^{(2)}(t)$ as a function of the scaled time $\lambda t$ with the same data as taken before. As it is seen from these figures, the Rabi-like oscillations in the temporal evolution of~$G^{(2)}(t)$ can be identified. Furthermore, one can see that the initial photon statistics of the cavity-field is only slightly distorted by the interaction with the atom. Figure 4(a) shows that when the influence of the gravitational field is negligible the cavity-field exhibits sub-Poissonian statistics in the course of interaction (Fig. 4(a)). With increasing $\vec k.\vec g$ the sub-Poissonian characteristic is suppressed and the cavity-field exhibits super-Poissonian statistics (Figs.4(b),4(c)). Also, numerical analysis reveals that by increasing the deformation parameter \textit{q}, the second-order correlation function shows chaotic-like oscillations and the super-Poissonian behaviour of the cavity-field in the presence of gravity is strengthened.

\subsection{Deformed quadrature squeezing of the cavity-field}
Squeezed states are shown by reduced quantum fluctuations in one quadrature
of the field at the expense of the increased fluctuations in the other quadrature. In order to investigate the quadrature squeezing of the \textit{f}-deformed cavity-field in the presence of homogenous gravitational field, we introduce the two slowly varying Hermitian
$f$-deformed quadrature components~$\hat{X}_{1A}$ and~$\hat{X}_{2A}$ defined, respectively, by

\begin{eqnarray}\label{46}
\hat{X}_{1A}(t)&\equiv &\frac{1}{2}(\hat{A} e^{i\nu t}+\hat{A}^{\dag} e^{-i\nu t}),\nonumber\\
\hat{X}_{2A}(t)&\equiv &\frac{1}{2i}(\hat{A} e^{i\nu t}-\hat{A}^{\dag} e^{-i\nu t}),
\end{eqnarray}
where the operators $\hat A$ and $\hat A^\dagger$are the \textit{f}-deformed annihilation and creation operators given by (\eqn{2}). In the limiting case $f(\hat n)=1$, the operators~$\hat{X}_{1A}$ and~$\hat{X}_{2A}$ reduce to the conventional (nondeformed) quadrature operators \cite{68}. These \textit{f}-deformed quadrature operators obey the commutation relation\begin{equation}\label{47}
\begin{array}{rcl}
[\hat{X}_{1A}(t),\hat{X}_{2A}(t)]=i/2((\hat{n}+1)f^2(\hat{n}+1)-\hat{n}f^2(\hat{n})),
\end{array}
\end{equation}
and, as result, the variances $<(\Delta\hat{X}_{jA}(t))^2>\equiv<(\hat{X}_{jA}(t))^2>-<(\hat{X}_{jA}(t))>^2$ satisfy the uncertainty relation
\begin{equation}\label{49}
\begin{array}{rcl}
 \left\langle {(\Delta \hat X_{1A}^{} (t))^2 } \right\rangle \left\langle {(\Delta \hat X_{2A}^{} (t))^2 } \right\rangle  \ge \frac{1}{{16}}\left( {\left\langle {(\hat n + 1)f^2 (\hat n + 1) - \hat nf^2 (\hat n)} \right\rangle } \right)^{{\rm  }2}.
\end{array}
\end{equation}
A state of the field is said to be squeezed when one of the \textit{f}-deformed quadrature components~$\hat{X}_{1A}$ and~$\hat{X}_{2A}$ satisfies the relation
\begin{equation}\label{48}
\begin{array}{rcl}
\Big<(\Delta\hat{X}_{jA}(t))^2\Big> <1/4(\Big<(\hat{n}+1)f^2(\hat{n}+1)-\hat{n}f^2(\hat{n})\Big>),~(j=1\,\,or\,\, 2)
\end{array}
\end{equation}
The degree of squeezing can be measured by the squeezing parameter $S_j(j=1,2)$ defined by
\begin{equation}\label{49.5}
\begin{array}{rcl}
S_j(t)\equiv 4<(\Delta\hat{X}_{jA}(t))^2>-(\Big<(\hat{n}+1)f^2(\hat{n}+1)-\hat{n}f^2(\hat{n})\Big>),
\end{array}
\end{equation}
which can be expressed in terms of the \textit{f}-deformed annihilation and creation operators of the field as follows

\begin{eqnarray}\label{50}
S_1(t)&=&2\xi(t)+2Re(\zeta(t))-4(Re(\eta(t)))^2 ,\nonumber\\
S_2(t)&=&2\xi(t)-2Re(\zeta(t))-4(Im(\eta(t)))^2,
\end{eqnarray}
where
\begin{equation}\label{51}
\begin{array}{rcl}
\xi(t)\equiv<\hat{A}^\dag \hat{A}>,\,\, \eta(t)\equiv<\hat{A}>e^{i\nu t},\,\, \zeta(t)\equiv<\hat{A^2}>e^{2i\nu t}.
\end{array}
\end{equation}
Then, the condition for squeezing in the \textit{f}-deformed quadrature component can be simply written as $S_j(t)<0$.

Now, we investigate the temporal behaviour of $S_2(t)$, which gives information on deformed quadrature squeezing of $\hat{X}_{2A}(t)$, when we consider the same data as used before. Numerical results are presented in Figs.5(a)-5(c), where we have plotted $S_2(t)$ versus $\lambda t$. Figure 5(a) shows that despite the fact that the cavity-field remains in the state quite close to the initial state, the \textit{f}-deformed quadrature components $\hat{X}_{2A}(t)$ exhibits squeezing in the course of time evolution when the influence of the gravitational field is negligible. However, in view of Figs. 5(b) and 5(c) we observe that the gravitational field leads to the disappearance of quadrature squeezing completely. Besides, numerical analysis shows that by increasing the nonlinearity of the model (increasing the value of the deformation parameter \textit{q}) the quadrature squeezing occurs in smaller amount for a short time in the very beginning of the interaction when the influence of gravity is negligible. Furthermore, with the increasing value of $\vec k.\vec g$ the quantum fluctuations in the quadrature component $\hat{X}_{2A}(t)$ increase considerably.
\section{Conclusions}

We proposed an \textit{f}-deformed formalism to study the quantum statistical properties of nonlinear JC model, including possible forms of nonlinearities of both the field and the intensity-dependent atom-field coupling, in the presence of a classical gravitational field. The work here extends previous studies in this context \cite{38,58}. To analyze the dynamical evolution of the atom-radiation system, we presented a quantum treatment of the internal and external dynamics of the atom based on an alternative \textit{f}-deformed $su(2)$ dynamical algebraic structure. By solving the Schr\"{o}dinger equation in the interaction picture, we found the evolving state of the system by which the influence of the classical gravity field on the dynamical behaviour of the atom-radiation system was explored. Whilst the model was quite general, we looked specifically at a special choice of the nonlinearity, i.e., \textit{q}-nonlinearity. Considering the cavity-field to be initially in a \textit{maths}-type \textit{q}-deformed coherent state and the two-level atom to be in a coherent superposition of its excited and ground states, we investigated the influence of gravity on the atomic population inversion, atomic dipole squeezing, atomic momentum diffusion, photon counting statistics, and  deformed quadrature squeezing of the radiation field. We found: (1) with the increasing value of the gravity-dependent parameter $\vec k.\vec g$ and nonlinearity (deformation) parameter \textit{q} the Rabi-like oscillations in the atomic population inversion are deteriorated. Also, the presence of the gravitational field causes more energy to be stored in the cavity-field while in the absence of gravitational influence, the deformation causes more energy to be stored in the atomic subsystem, (2) both the gravitational field and the nonlinearity associated with the \textit{f}-deformation cause the atomic dipole squeezing to be weakened, (3) both the gravitational influence and the \textit{f}-deformation cause the atom to experience larger light-induced forces during its interaction with the radiation field, (4) with increasing the parameters $\vec k.\vec g$ and \textit{ q} the sub-Poissonian behaviour of the cavity field is suppressed and it exhibits super-Poissonian statistics and (5) the deformed quadrature squeezing of the cavity-field decreases with increase of the gravity-dependent parameter $\vec k.\vec g$ and the deformation parameter \textit{q}. To make the model under consideration more realistic the atom decay and cavity-field decay should be taken into account. We hope to report on such issues in a forthcoming paper.
\section*{Acknowledgments}
The authors would like to express their gratitude to the referees,
whose valuable comments have improved the paper. They are
also grateful to the office of Graduate Studies of the University
of Isfahan for their support.
\vskip 2cm
\bibliographystyle{apsrev}

\vskip 16cm

\section*{Figure Captions}
\textbf{Figure1.} Time evolution of the atomic population inversion $W(t)$ versus the scaled time $\lambda t$. Here we have set $M = 10^{ - 26} kg,\alpha  = 2,\lambda  = 10^5 \,rad/s,k = 10^7 \,m^{ - 1}, g=10\, m.sec^{-2},\Delta _k  = 3 \times 10^7\, rad/s,\phi (\vec p) = \frac{1}{{\sqrt {2\pi } }}\exp ( - p^2 ), c_e(0)=c_g(0)=1/\sqrt{2},f(\hat n) = \sqrt {\frac{1}{{\hat n}}\frac{{1 - q^{\hat n} }}{{1 - q}}}$ and $q=1.04$ (1a) For~$\vec k.\vec g=0$ (1b) for~$\vec k.\vec g=2\times10^7\,\,sec^{-2}$(1c) for~$\vec k.\vec g=8\times10^7\,\,sec^{-2}$.\\
\\
\textbf{Figure2.} Time evolution of $F_y(t)$, corresponding to the squeezing of the component $\hat \sigma_y(t)$, versus the scaled time $\lambda t$, with the same corresponding data used in figure 1:(1a) For~$\vec k.\vec g=0$ (1b) for~$\vec k.\vec g=2\times10^7\,\,sec^{-2}$(1c) for~$\vec k.\vec g=8\times10^7\,\,sec^{-2}$.\\
\\
\textbf{Figure 3.} Time evolution of the atomic momentum diffusion $\Delta p(t)$ versus the scaled time $\lambda t$, with the same corresponding data used in figure 1:(1a) For~$\vec k.\vec g=0$ (1b) for~$\vec k.\vec g=2\times10^7\,\,sec^{-2}$(1c) for~$\vec k.\vec g=8\times10^7\,\,sec^{-2}$.\\
\\
\textbf{Figure 4.} Time evolution of the second-order correlation function $G^{2}(t)$ versus the scaled time $\lambda t$, with the same corresponding data used in figure 1:(1a) For~$\vec k.\vec g=0$ (1b) for~$\vec k.\vec g=2\times10^7\,\,sec^{-2}$(1c) for~$\vec k.\vec g=8\times10^7\,\,sec^{-2}$.\\
\\
\textbf{Figure 5.} Time evolution of the squeezing parameter $S_2(t)$, corresponding to the squeezing of the deformed quadrature component $X_{2A}$versus the scaled time $\lambda t$, with the same corresponding data used in figure 1:(1a) For~$\vec k.\vec g=0$ (1b) for~$\vec k.\vec g=2\times10^7\,\,sec^{-2}$(1c) for~$\vec k.\vec g=8\times10^7\,\,sec^{-2}$.
\vskip 6cm
\begin{figure}
    \begin{center}
    \includegraphics[width=2.4in]{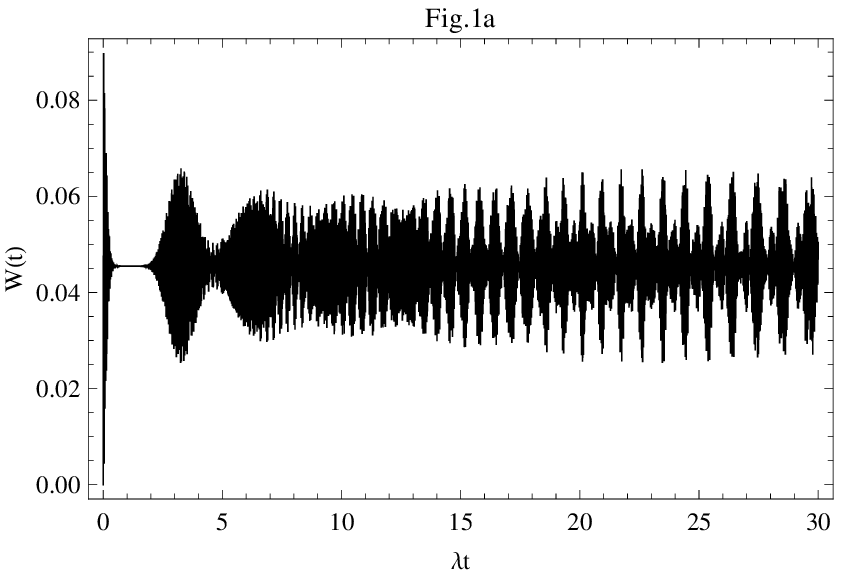}
    \includegraphics[width=2.4in]{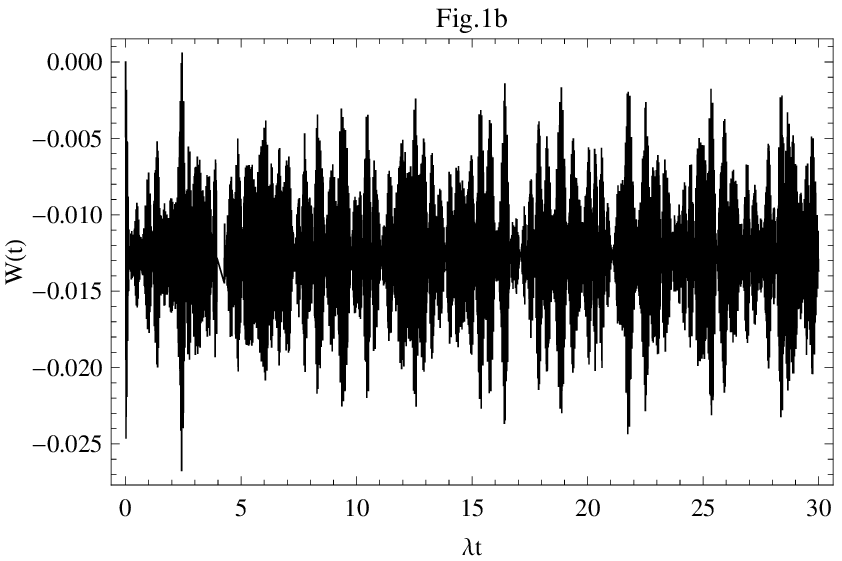}
    \includegraphics[width=2.4in]{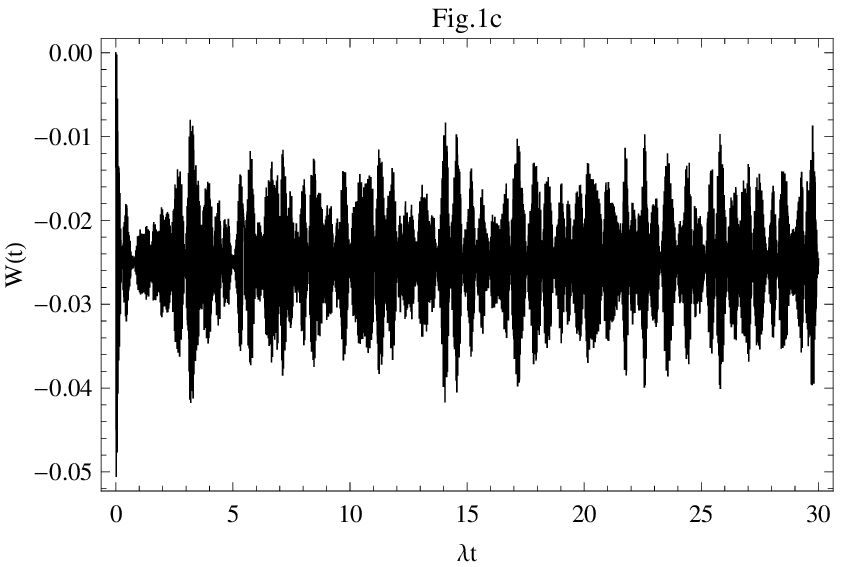}
   \caption{}
    \end{center}
\vskip 4cm

     \begin{center}
    \includegraphics[width=2.4in]{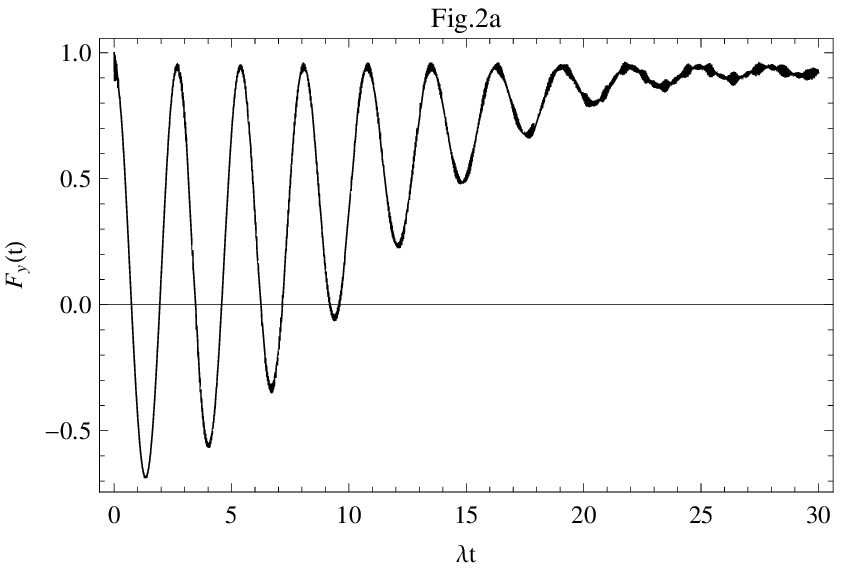}
    \includegraphics[width=2.4in]{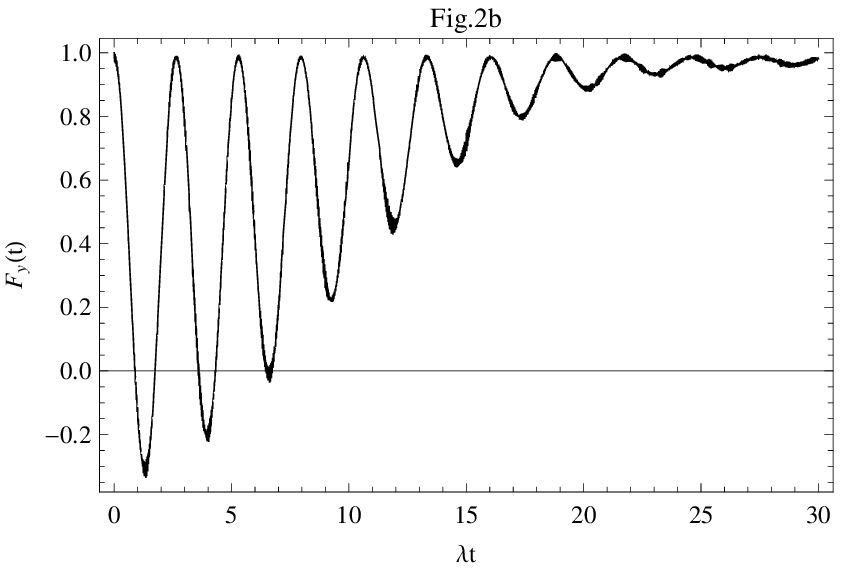}
    \includegraphics[width=2.4in]{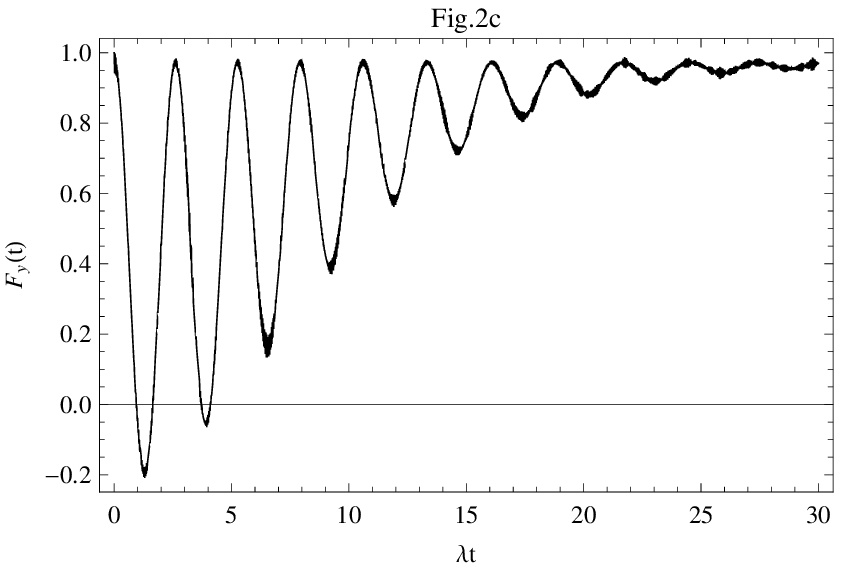}
\caption{}
    \end{center}
    \vskip 4cm

\end{figure}
\begin{figure}
    \begin{center}
    \includegraphics[width=2.4in]{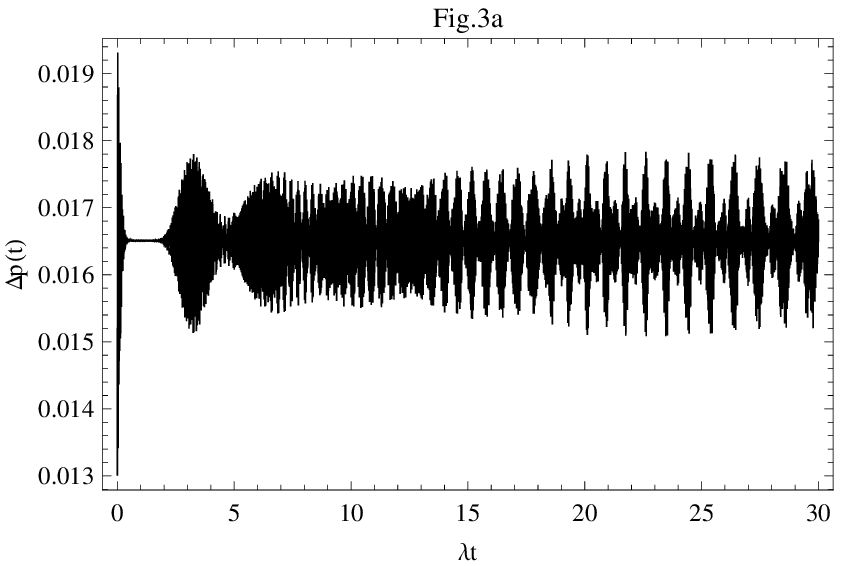}
    \includegraphics[width=2.4in]{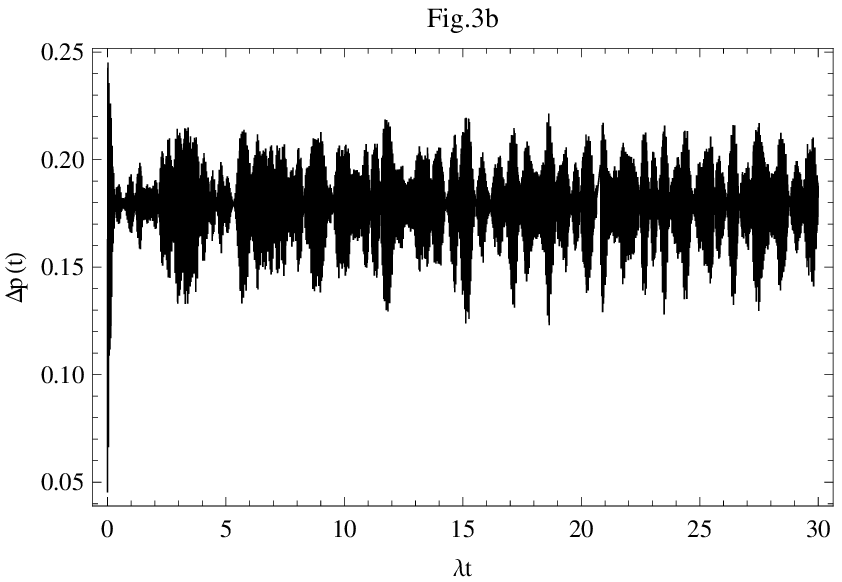}
    \includegraphics[width=2.4in]{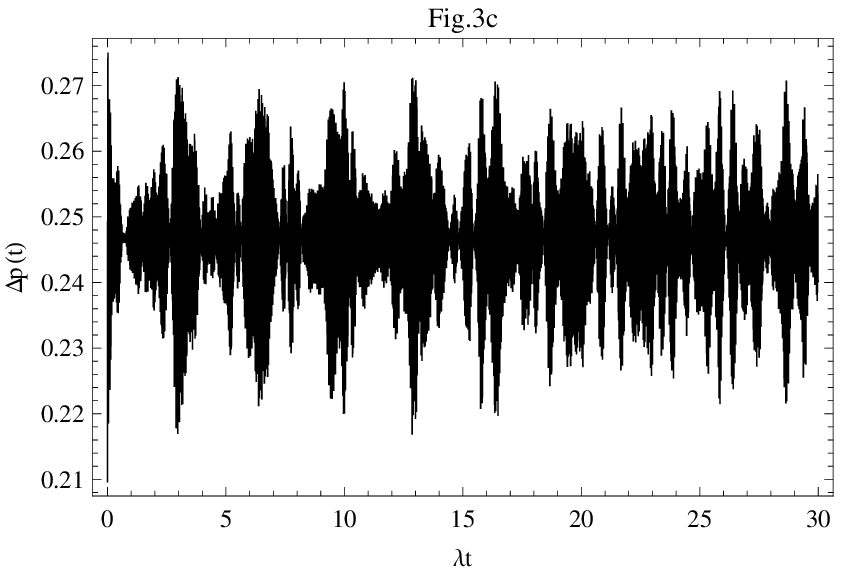}
   \caption{}
    \end{center}

\vskip 4cm

     \begin{center}
    \includegraphics[width=2.4in]{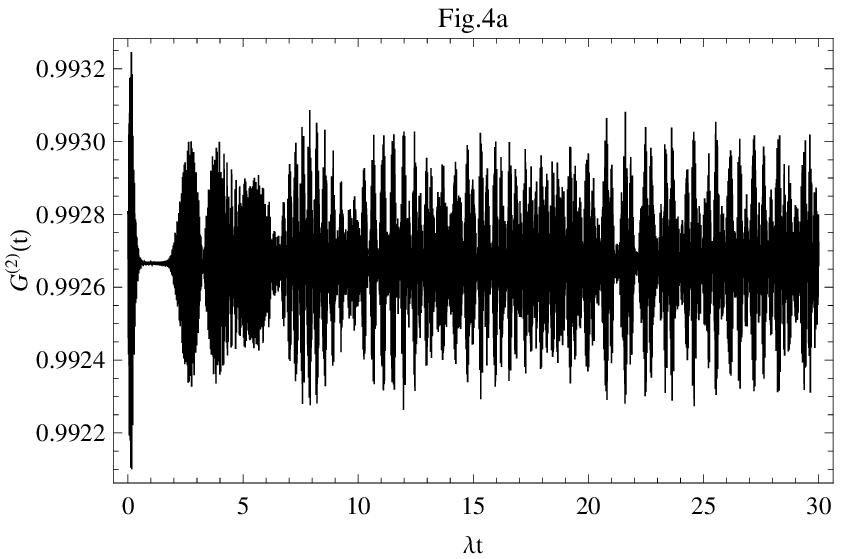}
    \includegraphics[width=2.4in]{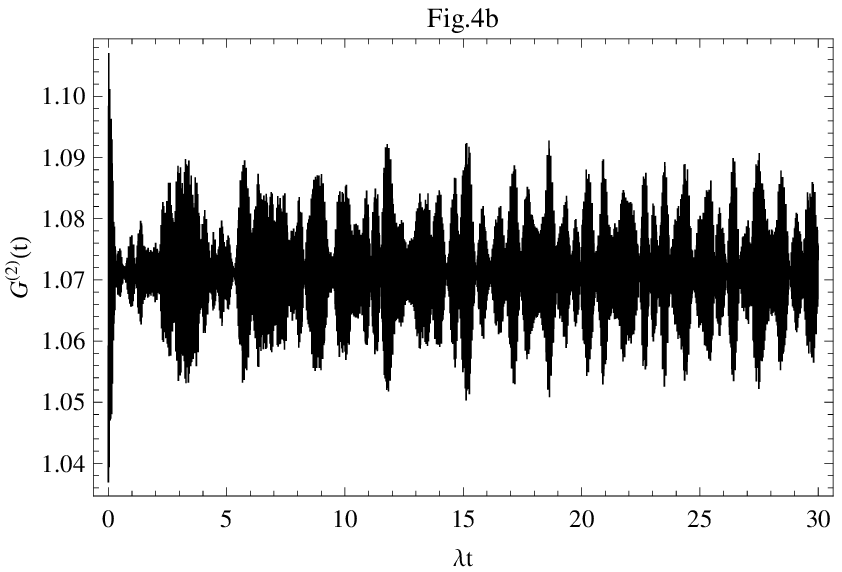}
    \includegraphics[width=2.4in]{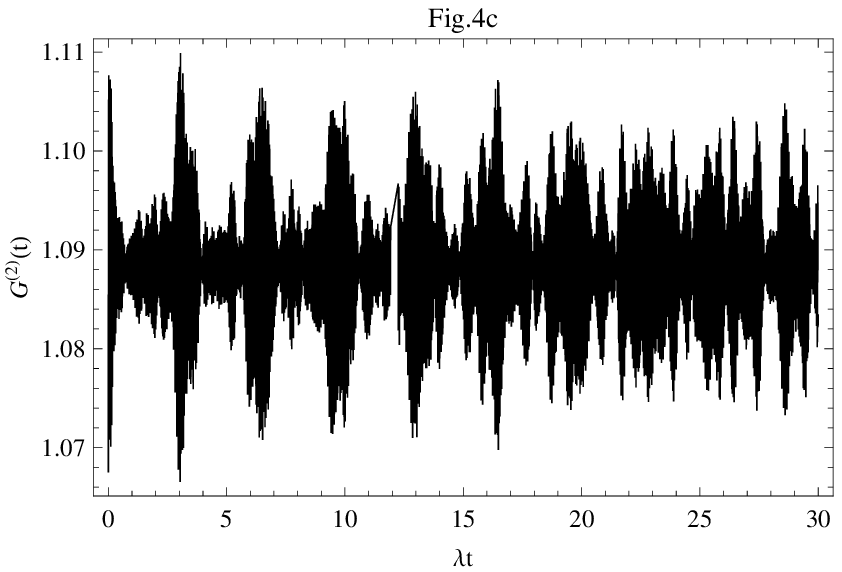}
   \caption{}
    \end{center}
\end{figure}
\vskip 4cm

\begin{figure}
    \begin{center}
    \includegraphics[width=2.4in]{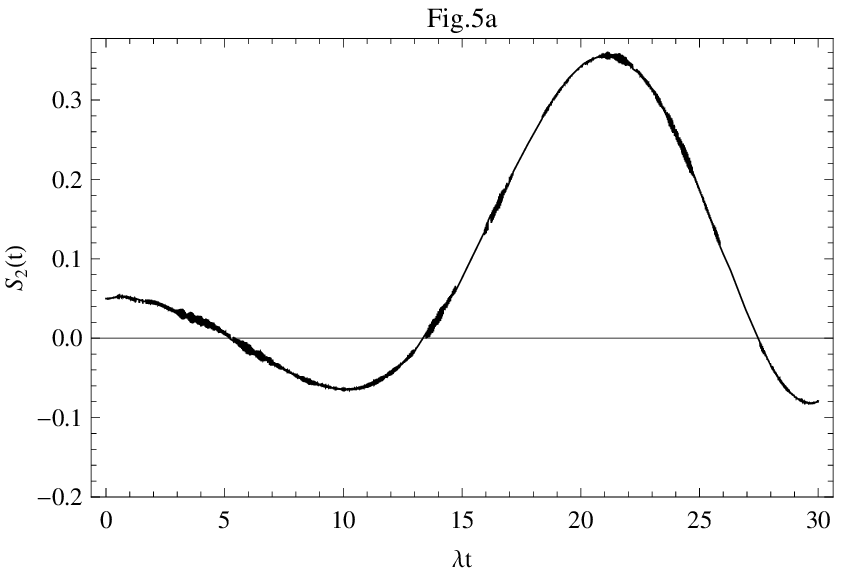}
    \includegraphics[width=2.4in]{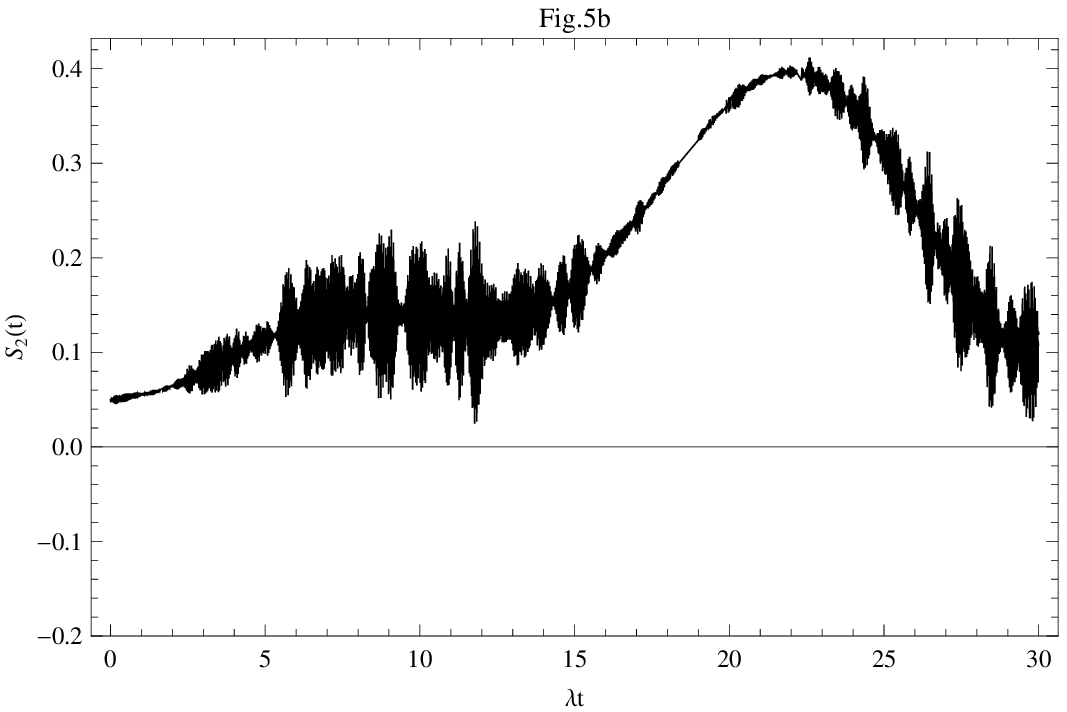}
    \includegraphics[width=2.4in]{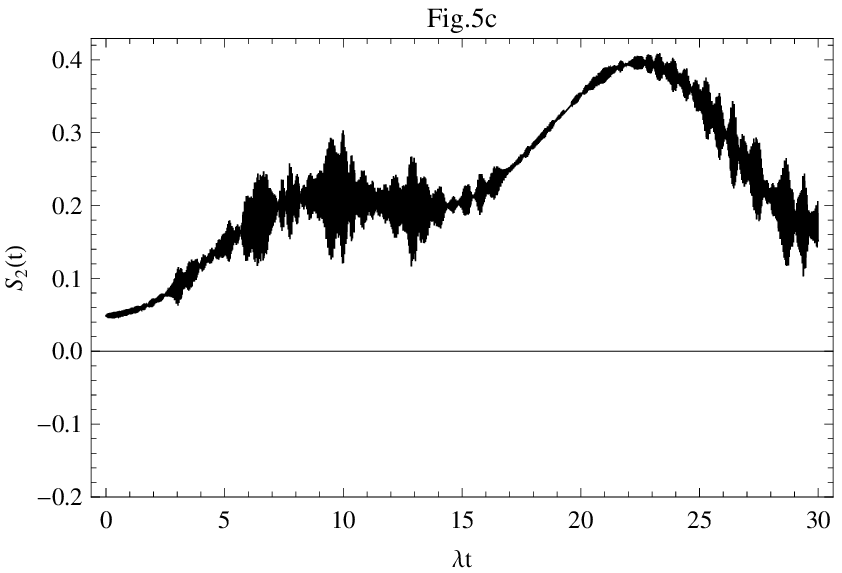}
   \caption{}
    \end{center}
     \end{figure}
\end{document}